\documentclass[prb,superscriptaddress,showpacs,twocolumn,floatfix]{revtex4}

\usepackage{amsmath}
\usepackage{braket}
\usepackage{graphicx}
\usepackage{epstopdf}

\begin{document}
\title{Magnetic detonation structure in crystals of molecular magnets}
\author{O. Jukimenko}
\affiliation {Department of Physics, Ume{\aa} University, SE-901\,87
Ume{\aa}, Sweden}

\author{M. Modestov}
\affiliation{Nordita, AlbaNova University Center, SE-106\,91
  Stockholm, Sweden}

\author{M. Marklund}
\affiliation {Department of Physics, Ume{\aa} University, SE-901\,87
Ume{\aa}, Sweden}
\affiliation{Department of Applied Physics, Chalmers University of Technology, SE-412 96 G\"{o}teborg, Sweden}

\author{V. Bychkov}
\affiliation {Department of Physics, Ume{\aa} University, SE-901\,87
Ume{\aa}, Sweden}

\begin{abstract}
Experimentally detected ultrafast spin-avalanches spreading in crystals of molecular (nano)magnets (Decelle et al., Phys. Rev. Lett. \textbf{102}, 027203 (2009)), Ref. [\onlinecite{Decelle-09}], have been recently explained in terms of magnetic detonation (Modestov et al., Phys. Rev. Lett. \textbf{107}, 207208 (2011)), Ref. [\onlinecite{Modestov-det}].
Here magnetic detonation structure is investigated by taking into account transport processes of the crystals such as thermal conduction and volume viscosity.  In contrast to the previously suggested model,  the transport processes result in  smooth profiles of the most important thermodynamical crystal parameters -- such as temperature, density and pressure -- all over the magnetic detonation front including the leading shock, which is one of the key regions of magnetic detonation.
In the case of zero volume viscosity, thermal conduction leads to an isothermal discontinuity instead of the shock, for which temperature is continuous while density and pressure experience jump.
\end{abstract}
\maketitle
\section{Introduction}

\begin{figure}
\centering
\includegraphics[width=3.3in]{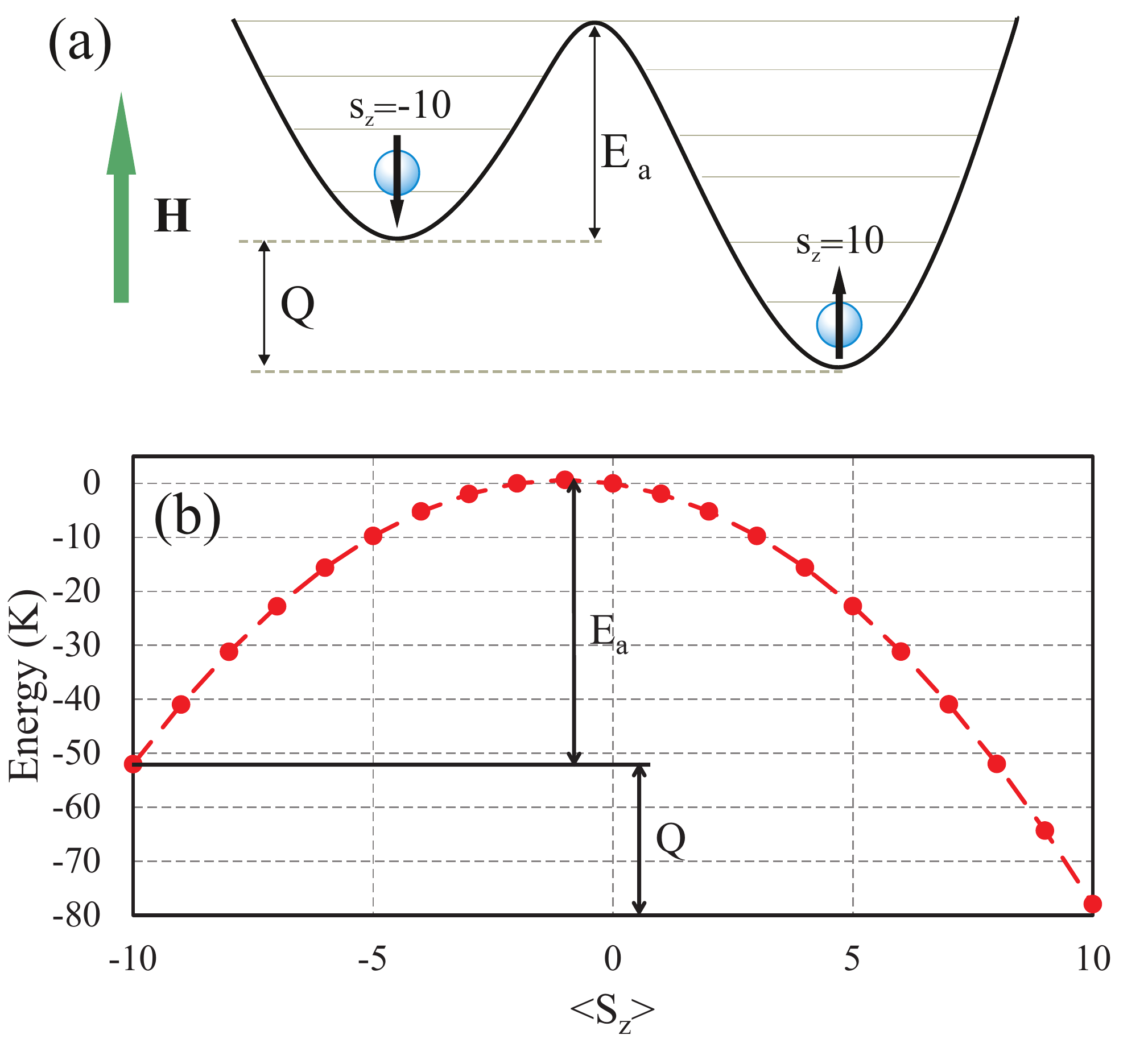}
\caption{\label{fig:energies_par}(Color online) (a) Schematic
  representation of the double-well structure of a nanomagnet.  (b)
  Energy levels for Mn$_{12}$-acetate in the external magnetic field
  $B_{z}=1\ \textrm{T}$. Axis $z$ is parallel to the easy axis of the
  crystal.  The energy barrier (activation energy) $E_{a}$ and the
  Zeeman energy $Q$ are indicated.}
\end{figure}

Presently there is much interest in molecular (nano) magnets with unique superparamagnetic properties, which may be used for quantum computing and memory storage. \cite{Bogani-Wernsdorfer-2008,Sarachik-2013,Suzuki-05,Hernandez-PRL-05,Subedi-2013}
A remarkable feature of nanomagnets is that, in contrast to classical magnets, these macromolecules with large effective molecular spin (e.g., $S=10$ for
Mn$_{12}$-acetate) can keep their spin-orientation upon the reversal of the
external magnetic field.\cite{mag:sessoli93,mag:villain94} Because of the strong molecular anisotropy, spin of a nanomagnet is
directed preferentially along the so-called
\emph{easy} axis, and it leads to a considerable energy barrier
between the spin-up and spin-down states.
At low temperatures, in a
 magnetic field directed along the easy axis, the states
with spin along the field and against the field become stable and
metastable, respectively. The energy difference between the two states
is determined by the Zeeman energy, $Q$, as illustrated in
Fig.~\ref{fig:energies_par}, with the energy barrier designated by
$E_{a}$. The barrier hinders spontaneous quantum
tunneling from the metastable to stable state  at low
temperatures,
\cite{mag:Friedman96,mag:thomas96,mag:chudnovsky98,mag:wernsdorfer01,mag:gatteschi03}
so that  fast spin-flipping requires help from outside.

For nanomagnets composing a crystal, relatively fast spin-flipping of one particular molecule may be induced by energy supplied by its' neighbors. When all or most of the molecules of a crystal are initially in the metastable state, then 
local heating by an external source  may trigger local spin-flipping, with Zeeman energy released in the heated region and transported to the next layer of the crystal.\cite{Sarachik-2013,Suzuki-05,Hernandez-PRL-05,Subedi-2013,Garanin-Chudnovsky-2007,Modestov-2011,McHugh-09,Dion-2013,Jukimenko-2014} The heat facilitates spin-flipping in the next layer, and so on, so that  the process spreads in a crystal as a thin self-supporting magnetization front -- spin-avalanche --  well-localized spatially. Usually, energy in spin-avalanches is transported from one crystal layer to another by means of thermal conduction, and hence a spin-avalanche propagates at moderate speed,  $\sim1-10$ m/s.\cite{Sarachik-2013,Suzuki-05,Hernandez-PRL-05,Subedi-2013} Due to striking similarity of such avalanches to slow combustion flame, \emph{deflagration},  avalanches of this type are typically called "magnetic deflagration".

In contrast to the slow magnetic deflagration studied in the absolute majority of works on the subject,\cite{Sarachik-2013,Suzuki-05,Hernandez-PRL-05,Subedi-2013,Garanin-Chudnovsky-2007,Modestov-2011,McHugh-09,Dion-2013,Jukimenko-2014}
recent experiments of Ref. [\onlinecite{Decelle-09}] detected ultrafast spin-avalanches propagating at speed comparable to the sound speed in the crystals, $\approx 2000$ m/s. The theory presented in Ref. [\onlinecite{Modestov-det}] has explained the ultrafast spin-avalanches in terms of "magnetic \emph{detonation}" and investigated the key properties of the process. In particular, it has been demonstrated that magnetic detonation belongs to the type of weak detonations and propagates with speed only slightly exceeding the sound speed $\approx 2000$ m/s. Paper [\onlinecite{Modestov-det}] has also studied structure of magnetic detonation within the traditional combustion model of a detonation front consisting of an infinitely thin leading shock and a zone of energy release of finite thickness.\cite{Law-book} Such a model has been originally developed for gaseous detonations, which are quite strong and propagate with speed exceeding the sound speed in gases by an order of magnitude. However, it is not rigorous applying such a model to  magnetic detonation, which is extremely weak by combustion scales (although quite strong when compared to magnetic deflagration). In contrast to strong shocks in gases, which are infinitely thin from the hydrodynamic point of view, a weak shock exhibits a continuous structure controlled by transport processes such as thermal conduction and/or viscosity.\cite{LL_VI}  The same property should be naturally expected for magnetic detonation.

The purpose of the present work is to provide accurate description of magnetic detonation structure in crystals of nanomagnets by taking into account thermal conduction and volume viscosity.
Here we show that, in contrast to the previously suggested model of Ref. [\onlinecite{Modestov-det}],  the transport processes result in  smooth profiles of the most important thermodynamical crystal parameters -- such as temperature, density and pressure -- all over the magnetic detonation front including the leading weak shock, which is one of the key regions of magnetic detonation.
In the case of zero volume viscosity, however, thermal conduction leads to an isothermal discontinuity instead of the shock, for which temperature is continuous while density and pressure experience jump.

The present paper is organized as follows. In Sec. II we overview basic features of gaseous combustion detonation needed for proper understanding of magnetic detonation in crystals of nanomagnets. In Sec. III we present basic equations describing magnetic detonation, develop the analytical theory for the most important detonation parameters and study magnetic detonation structure controlled by thermal conduction only assuming zero volume viscosity -- such a structure involves isothermal discontinuity instead of a shock employed in Ref. [\onlinecite{Modestov-det}]. In Sec. IV we demonstrate dramatic modifications of the magnetic detonation structure due to volume viscosity.


\section{Basic features of gaseous combustion detonation}

\begin{figure}
\includegraphics[width=3.4in]{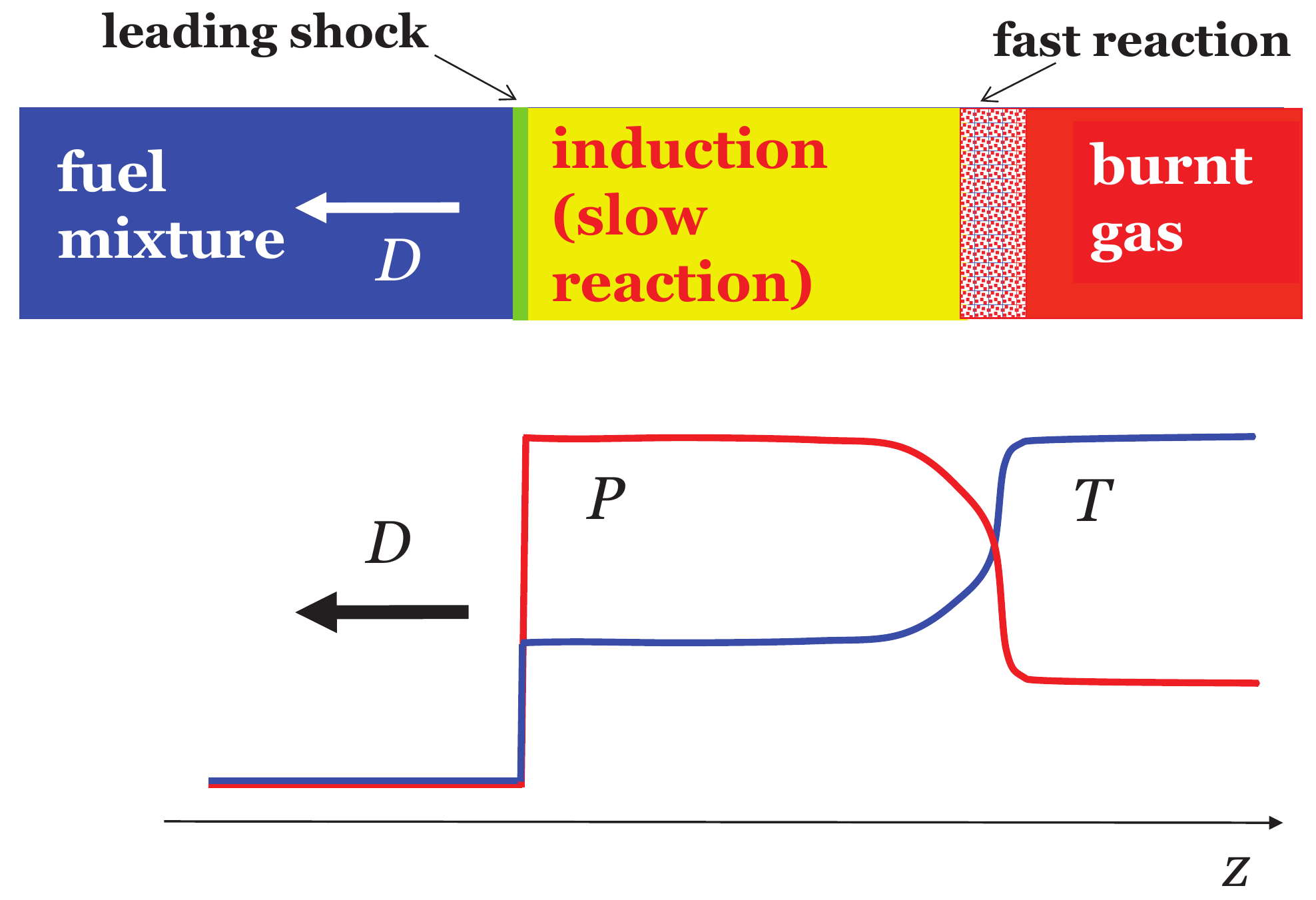}
\caption{\label{fig:Det-structure_schematic}
Schematic of strong gaseous detonation with characteristic profiles of temperature and pressure.}
\end{figure}

In this section we remind briefly the most important features  and the methods of investigation of traditional gaseous combustion detonations, in order to highlight the similarity and difference between the combustion and magnetic detonations. By definition, detonation is a fast supersonic combustion regime, for which preheating of the cold fuel mixture happens due to the leading shock, see Fig.~\ref{fig:Det-structure_schematic}. In turn, shock propagation without decay in combustion detonation is supported by energy release in chemical reactions in the active reaction zone. Typically for combustion, the activation energy of the reactions is quite high, so that reactions develop relatively slow just after the leading shock in the so-called induction zone (still, much faster than in the fuel mixture). It requires certain (induction) time for self-acceleration of chemical reactions in a gas parcel after passing the shock, after which the chemical reactions become fast and convert the fuel mixture into the burning products with release of a large amount of energy and strong  expansion of the burning gas.

Here we are interested in a planar stationary one-dimensional detonation propagating with constant \emph{supersonic} speed $D$ in a uniform gaseous mixture; we take the detonation front propagating along the z-axis in the negative direction. By adopting the reference frame of the stationary detonation front, we obtain the fuel mixture moving with velocity  $u_{0} = D$ towards the leading shock of zero thickness; the initial density, pressure and temperature are designated by $\rho_{0}$, $P_{0}$ and $T_{0}$. The fuel mixture is compressed in the shock (label "$s$") with strong increase of density, pressure and temperature to $\rho_{s}>\rho_{0}$, $P_{s}>P_{0}$, $T_{s}>T_{0}$, and with drop of the gas velocity to a new \emph{subsonic} value $u_{s}<D$. Temperature increase initiates combustion reactions with release of chemical energy, hence leading to even stronger temperature increase until it reaches the final maximal value in the detonation products (label "$d$"), $T_{d}>T_{s}$. At the same time, pressure and density decrease in the reaction region,  $\rho_{d}<\rho_{s}$, $P_{d}<P_{s}$, as illustrated schematically on Fig.~\ref{fig:Det-structure_schematic}.

Modifications of the gas parameters in a detonation front are described by the hydrodynamic
 laws of mass, momentum and energy conservation expressed in terms of fluxes\cite{LL_VI}
\begin{equation}
\label{eq:mass1}
\rho u=const_{1}=\rho_{0} D,
\end{equation}
\begin{equation}
\label{eq:momentum1}
P+\rho u^2=const_{2}= P_{0}+\rho_{0} D^2,
\end{equation}
\begin{equation}
\label{eq:energy1}
\rho u \left( h+\frac{u^{2}}{2}\right)=const_{3}=\rho_{0} D \left( h_{0}+\frac{D^{2}}{2}\right),
\end{equation}
where $h$ is the gas enthalpy
\begin{equation}
\label{eq:enthalpy}
h=\frac{\gamma P}{(\gamma-1) \rho}+a Q,
\end{equation}
$\gamma$ is the adiabatic exponent, $Q$ is the  chemical energy stored per  unit mass, $a$ is  fraction of the unburned fuel mixture, which changes from $a_{0}=a_{s}=1$ in the fresh and shocked gas to $a_{d}=0$ in the detonation products. The hydrodynamic equations Eqs. (\ref{eq:mass1})-(\ref{eq:energy1})
are complemented by the equation of state, which is taken for gaseous detonations in the form of the ideal gas law
\begin{equation}
P = \frac{\rho}{m} R T,
\label{eq:idgas}
\end{equation}
where $R$ is the ideal gas constant and $m$ is the molar mass of the gas (for simplicity we assume here that the molar mass does not change in the combustion process). Traditionally, the theory of shock waves and detonations employs volume per unit mass $V\equiv 1/\rho$ for analyzing the process.\cite{LL_VI}

\begin{figure}
\includegraphics[width=3.1in]{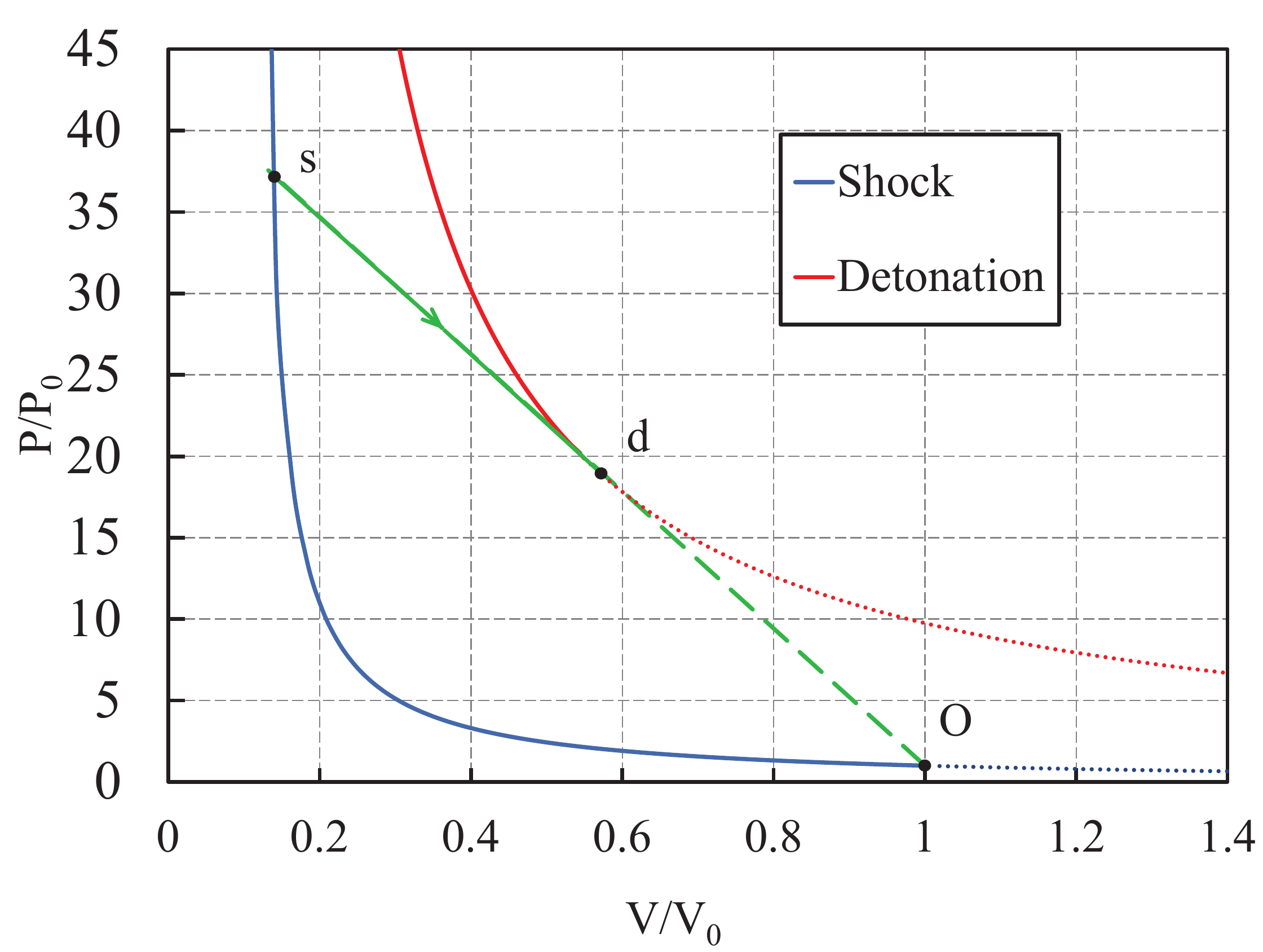}
\caption{\label{fig:Hugoniot_1}The pressure-volume diagram for shock (blue curve) and detonation (red curve) in a gas mixture. The scaled energy release is $(\gamma-1) Q/ \gamma P_{0}  V_{0} = 9$, with the initial volume per unit mass  $V_0=6.33\ \mathrm{m}^3/\mathrm{kg}$ and initial pressure $P_0=1.33\cdot10^4$ Pa. The green straight line shows transition from the shock point $s$  to the combustion products $d$ in the CJ detonation.}
\end{figure}

Modifications of the gas parameters in a leading shock are described by Eqs. (\ref{eq:mass1})-(\ref{eq:energy1}) with zero energy release in the reaction within the discontinuous shock, $a_{s}=1$. Then Eqs. (\ref{eq:mass1})-(\ref{eq:energy1}) may be reduced to the so-called Hugoniot equation, which specifies all possible finite states of the shocked gas $P_{s}, V_{s}$ for a fixed initial state $P_{0}, V_{0}$; the result is demonstrated  by the blue solid curve in Fig. \ref{fig:Hugoniot_1} for the initial state corresponding to the ideal gas  at initial volume per unit mass $V_0=6.33 \ \mathrm{m}^3/\mathrm{kg}$ and  pressure $P_0=1.33\cdot10^4$ Pa. For any particular final state $s$ at the shock, equations (\ref{eq:mass1}), (\ref{eq:momentum1}) relate the slope of the green straight line ($sO$) in Fig. \ref{fig:Hugoniot_1} to the front propagation speed $D$ as\cite{LL_VI}
\begin{equation}
\label{slope2}
\rho_{0}^{2}D^{2}=\frac{P_s-P_0}{V_0-V_s}.
\end{equation}

In order to obtain the final state of the detonation products we solve Eqs. (\ref{eq:mass1})-(\ref{eq:energy1}) for the case of complete burning  $a_{d}=0$; the result is plotted by the solid red curve in Fig. \ref{fig:Hugoniot_1} for the scaled energy release $(\gamma-1) Q/ (\gamma P_{0}  V_{0}) = 9$ corresponding to the acetylene-air combustion.\cite{Ott-2003} The red curve (for detonation) corresponds to higher pressure for the same volume than the blue curve (for a shock) due to the energy release in a detonation front. Still, unlike a shock, detonation speed is not a free parameter, but it is determined by particular boundary conditions. For example, the case of a freely propagating detonation corresponds to the so-called Chapman-Jouguet (CJ) detonation regime, with the detonation products moving away from the front at local sound speed.\cite{LL_VI} The CJ detonation propagates with the smallest speed possible for detonations for a fixed energy release, and indicated by the green tangent line in Fig. \ref{fig:Hugoniot_1}. Thus, on the pressure-volume diagram, the CJ detonation structure corresponds first to the jump from the  point $O$ (the initial state) to the point $s$ at the shock, and then the reaction development behind the shock along the green straight line from the point $s$ to the point $d$ (detonation products). It may be shown that the CJ detonation speed is determined by the energy release in the combustion process.\cite{LL_VI}

\begin{figure}
\includegraphics[width=3.4in]{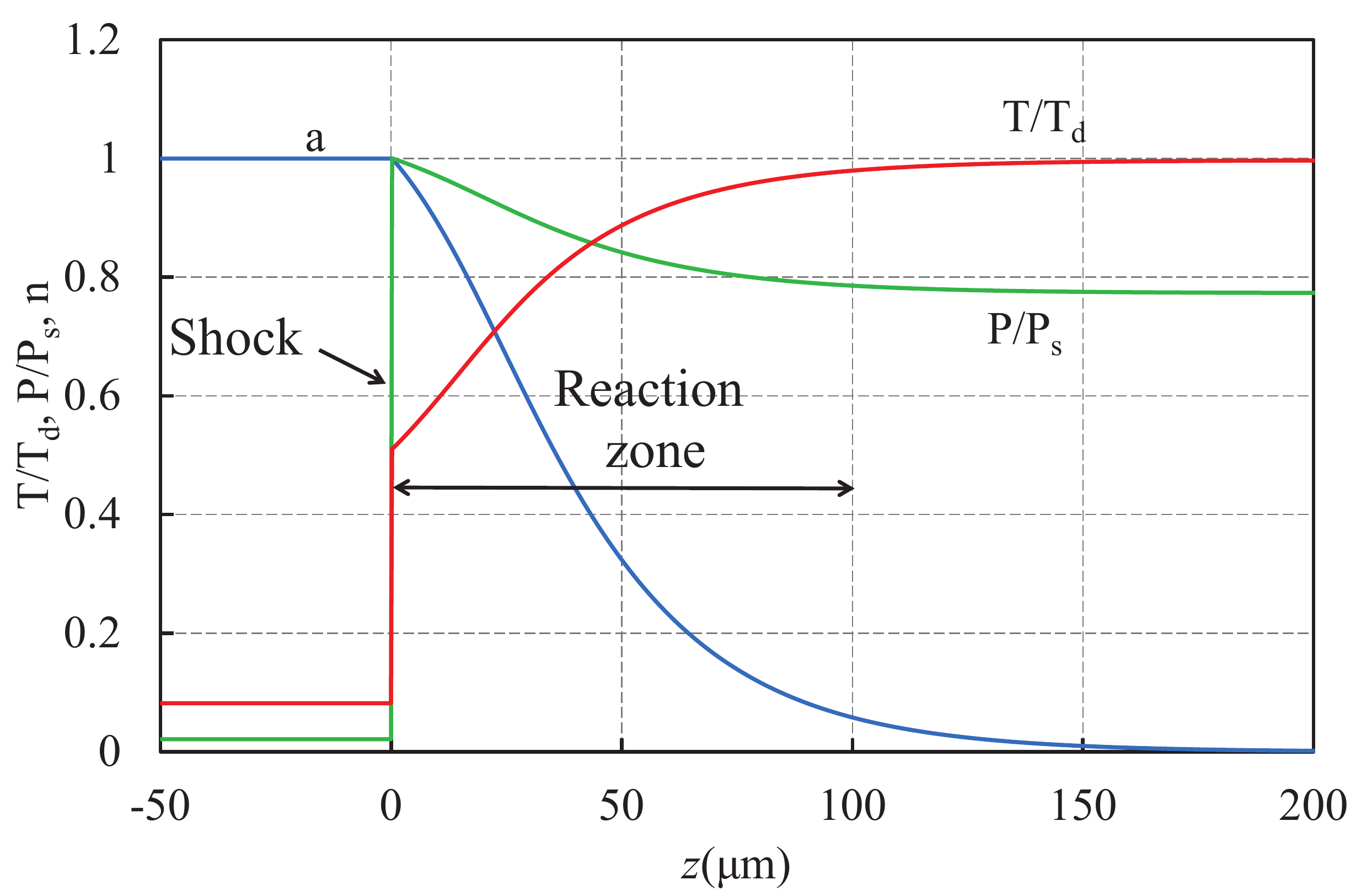}
\caption{\label{fig:Profile_id}Characteristic detonation structure in a gaseous mixture with  parameters representing acetylene-air combustion. The scaling is performed by using pressure at the shock $P_{s}=3.7\,\mathrm{MPa}$ and the temperature in the detonation products $T_{d}=2988\,\mathrm{K}$; other computational parameters are specified in the text. }
\end{figure}

The internal structure of the reaction zone in the detonation front (transition from the point $s$ to the point $d$ in Fig. \ref{fig:Hugoniot_1}) is specified by reaction kinetics of a particular fuel mixture. Here  we illustrate the detonation structure for the simplified model of acetylene-air combustion described by a single one-step  Arrhenius reaction\cite{Ott-2003}
\begin{equation}
\label{burning_1}
\frac{\partial a}{\partial t}=-K \rho a \exp(-E_{a}/T),
\end{equation}
where $K=10^{9} \,\textrm{m}^{3}/\textrm{kg}\,\textrm{s}$ is a pre-exponential factor, $E_{a}/T_{0}=29.3$ is the scaled activation energy.
Then the stationary detonation structure is determined by Eq. (\ref{burning_1}) rewritten in the reference frame of the front as
\begin{equation}
\label{burning_2}
u\frac{\partial a}{\partial z}=K \rho a \exp(-E_{a}/T).
\end{equation}
By solving Eq. (\ref{burning_2}) together with  Eqs. (\ref{eq:mass1})-(\ref{eq:energy1}) we find the detonation front structure as shown in Fig. \ref{fig:Profile_id}. For obtaining the plots we have used the scaled energy release $(\gamma-1) Q/ (\gamma P_{0}  V_{0}) = 9$, with the initial gas pressure $P_0=0.627\,\mathrm{MPa}$ and temperature $T_0=293\,\mathrm{K}$. Then by using the conservation laws Eqs. (\ref{eq:mass1})-(\ref{eq:energy1}) one can find the maximal pressure in the detonation front achieved at the shock as $P_{s}=3.7\,\mathrm{MPa}$ and maximal temperature achieved in the detonation products $T_{d}=2988\,\mathrm{K}$; these two values have been used for scaling in Fig. \ref{fig:Profile_id}. As we can see, Fig. \ref{fig:Profile_id} presents the detonation structure  similar to the  qualitative sketch of Fig. \ref{fig:Det-structure_schematic}, although the induction length is exhibited  not so strongly by the acetylene-air detonations.

\section{Magnetic detonation structure due to heat transfer}
\subsection{Basic equations}
The theory of combustion detonation summarized briefly in the previous section provides the clue for describing the magnetic detonation in crystals of nanomagnets; conceptually the same model has been employed for the analysis of magnetic detonation in Ref. [\onlinecite{Modestov-det}]. Still, here we stress important difference in some combustion and magnetic detonation properties, which is taken into account below. First, magnetic detonation in crystals of nanomagnets is extremely weak, so that it propagates with speed only slightly exceeding the sound speed $c_{0}$ in the crystal, $D \approx c_{0}$. For that reason, in stark contrast to combustion detonations, the leading shock in magnetic detonation is not a discontinuity, but a smooth transition region with structure determined by  transport processes, namely, thermal conduction $\kappa$ and volume viscosity $\eta$, which requires proper modifications of the conservation laws Eqs. (\ref{eq:mass1})-(\ref{eq:energy1}) at the detonation front.

So far the theoretical models of spin-avalanches in crystals of nanomagnets have taken into account only thermal conduction and neglected volume viscosity;\cite{Garanin-Chudnovsky-2007,Modestov-2011,McHugh-09,Dion-2013}  in the present section we use the same approach and consider the influence of thermal conduction only. Volume viscosity will be taken into account in the next section.
Although we deal with solid state, propagation of shocks and detonations in crystals of nanomagnets is also described by the hydrodynamic conservation laws of mass, momentum and energy similar to  Eqs. (\ref{eq:mass1})-(\ref{eq:energy1}), see Refs. [\onlinecite{Zeldovich-02,Modestov-det}]
\begin{equation}
\label{eq:mass1h}
\rho u=\rho_0 D,
\end{equation}
\begin{equation}
\label{eq:momentum1h}
P+\rho u^2=P_0+\rho_0 D^2,
\end{equation}
\begin{align}
\label{eq:energy1h}
\rho u &\left( \varepsilon+Qa+\frac{P}{\rho}+\frac{u^{2}}{2}\right)\nonumber-\kappa  \frac{dT}{dz}= \\
&\rho_0 D \left( \varepsilon_0+Q+\frac{P_0}{\rho_0}+\frac{D^{2}}{2}\right),
\end{align}
where $\varepsilon$ is the thermal energy per molecule, $Q$ stands for Zeeman energy and $a$ is the fraction of molecules in the metastable state.
Zeeman energy $Q$, together with the activation energy $E_a$, are determined by the applied magnetic field, and can be obtained by using Hamiltonian for Mn$_{12}$-acetate molecule \cite{mag:delbarco05}
\begin{equation}
\label{eq:Hamil}
\mathcal{H}=-D S_z^2-g \mu_B B_{z} S_z,
\end{equation}
where $B_{z}$ is the magnetic field, $g=1.94$ is the gyromagnetic factor, $\mu_B$ is the Bohr magneton, and $D=0.65 \mathrm{K}$ is the magnetic anisotropy constant.
 We assume that the field is applied along the easy axis (z-axis). Zeeman energy is found as the energy difference between the stable, $S_z=10$, and metastable, $S_z=-10$, states of the nanomagnet as illustrated in Fig.~\ref{fig:energies_par}
\begin{equation}
\label{eq:Zeeman}
Q= 2 g \mu_B B_{z} S\frac{R}{M}.
\end{equation}
Zeeman energy increases linearly with the magnetic field.
Activation energy is determined by the energy barrier  between stable and metastable states, see Fig.~\ref{fig:energies_par}, and for nanomagnets it may be found as
\begin{equation}
\label{eq:Ea}
E_a= D S^2-g \mu_B B_{z} S+\frac{g^2}{4 D} \mu_B^2 B_{z} ^2;
\end{equation}
 the activation energy is provided in temperature units. Dependence of Zeeman and activation energies on the applied magnetic field is presented in Fig.~\ref{fig:Q,Ea}
for  Mn$_{12}$-acetate. As we can see, the activation energy decreases with the magnetic field and turns to zero at $B_{z}\approx 10 \textrm{T}$; at higher fields the potential barrier disappears and the transition from the state $S_z=-10$ to the state $S_z=10$ is not hindered any more.
\begin{figure}
\includegraphics[width=3.4in]{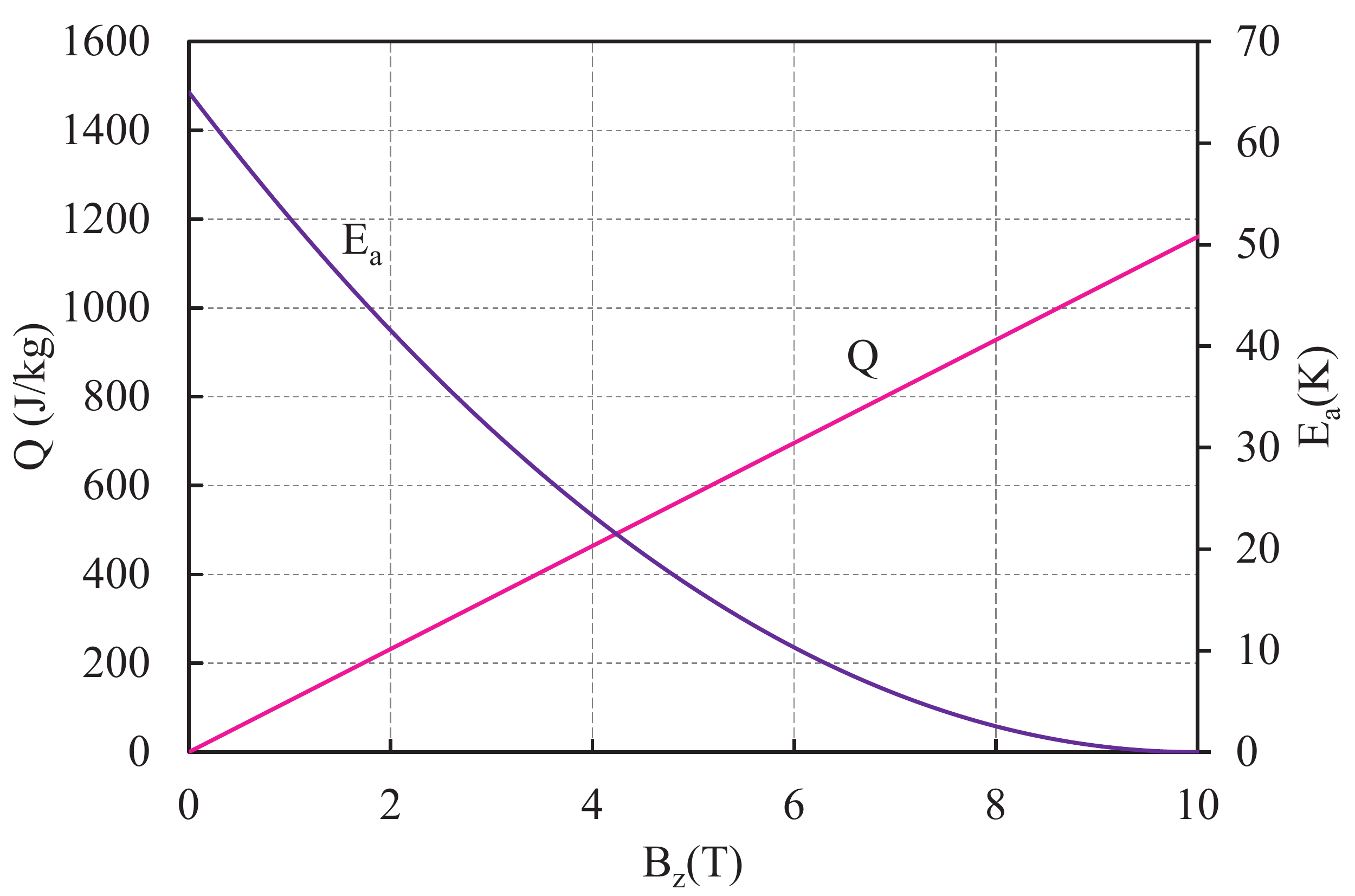}
\caption{\label{fig:Q,Ea} The Zeeman energy $Q$ and activation $E_a$, both in temperature units, vs the external magnetic field $B_{z}$.}
\end{figure}

It is convenient to define new dimensionless value $r\equiv \rho/\rho_0 = V_{0}/V$, which represents possible compression of the crystal in the process of magnetic detonation. 
Weak shock and detonation cause only elastic deformations, and the equation of state for the crystals of nanomagnets may be written as a combination of elastic and thermal components as\cite{Zeldovich-02,Modestov-det}
\begin{equation}
\label{eq:presure1}
\frac{P}{\rho_{0}}=\frac{c_{0}^{2}}{n}\left(r^{n}-1\right)+\frac{R A\Gamma k_{B}T^{\alpha +1}r}{M \left( \alpha +1\right)\Theta_{D}^{\alpha}},
\end{equation}
where $M$ is the molecular mass with $M=1868$ g/mol for Mn$_{12}$-acetate, see Ref. [\onlinecite{Sarachik-2013}]. Similar to Ref. [\onlinecite{Modestov-det}] we take
$n=4$,   the Gruneisen coefficient $\Gamma=2$,  the problem dimension $\alpha=3$, and the Debye temperature $\Theta_{D}=38$ K. The coefficient $A=12\pi^{4}/5$ corresponds to the Debye crystal lattice, and $c_0\approx2000\,\mathrm{m/s}$ is sound speed in the crystal.  Thermodynamic energy per one molecule of the crystal is given by\cite{Zeldovich-02,Modestov-det}
\begin{equation}
\label{eq:st_energy1}
\varepsilon=\frac{c_{0}^{2}}{n}\left(\frac{r^{n-1}-1}{n-1}+\frac{1}{r}-1\right)
+\frac{R AT^{\alpha+1}}{M \left(\alpha +1\right)\Theta_{D}^{\alpha}}.
\end{equation}
Equations (\ref{eq:mass1h})-(\ref{eq:energy1h}) provide a complete system for describing magnetic detonation parameters in  crystals of nanomagnets with  thermal conduction taken into account.

\begin{figure}
\includegraphics[width=3.4in]{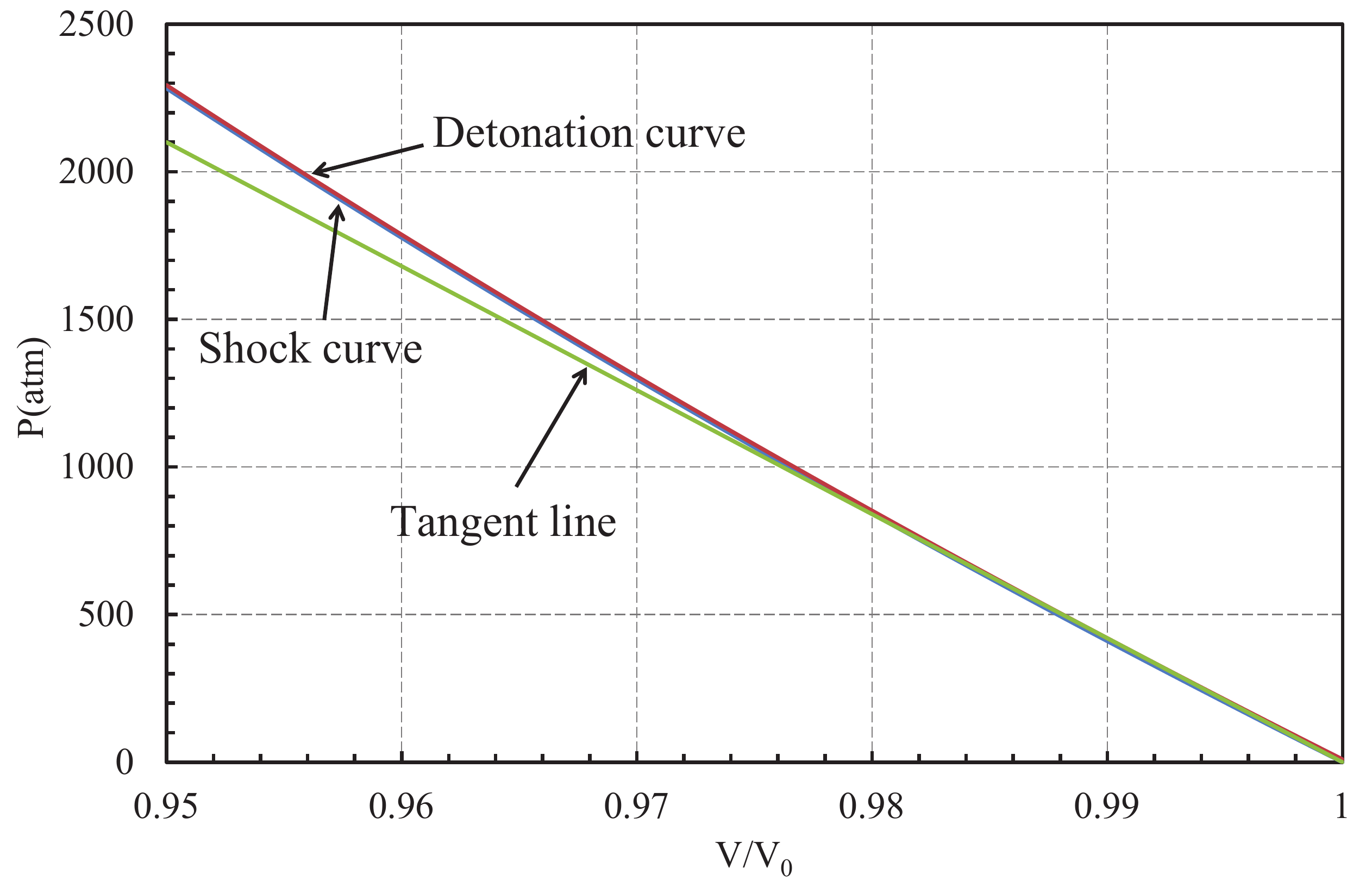}
\caption{\label{fig:PV_visc0} The pressure-volume diagram for the shock (blue) and magnetic detonation (red) in a Mn$_{12}$-acetate crystal. The green line represents the tangent line from the initial state to the CJ regime; the magnetic field is $B_{z}=4 \textrm{T}$.}
\end{figure}

\begin{figure}
\includegraphics[width=3.4in]{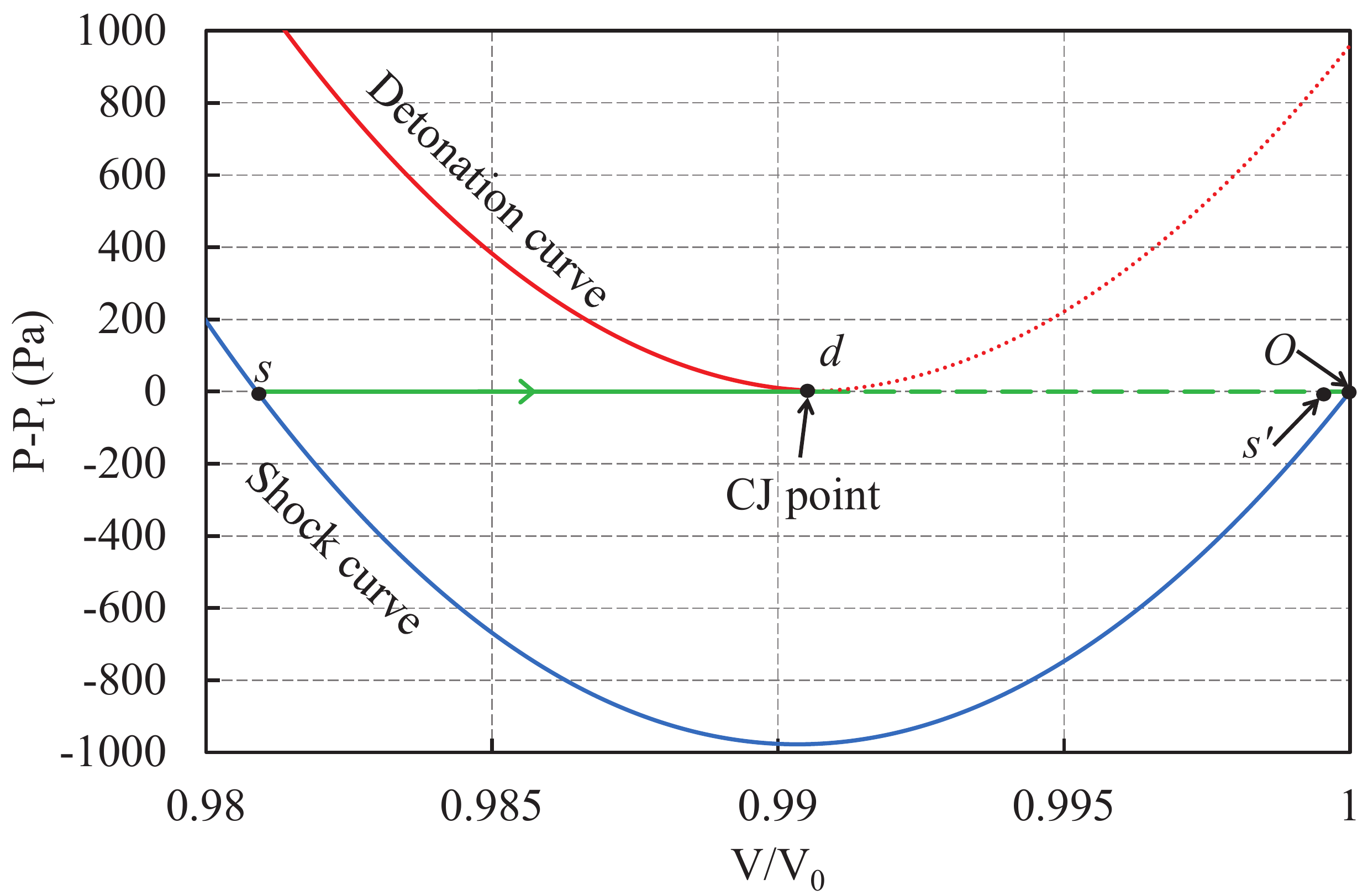}
\caption{The modified pressure-volume diagram for $P-P_{t}$ for the shock (blue) and magnetic detonation (red) in a Mn$_{12}$-acetate crystal. The reference pressure $P_{t}$ of the green line corresponds to the tangent line Eq. (\ref{eq:tang}) from the initial state to the CJ regime; the magnetic field is $B_{z}=4 \textrm{T}$.}
\label{fig:PV_kappa}
\end{figure}

Solution to Eqs. (\ref{eq:mass1h})-(\ref{eq:energy1h}) is presented in Fig.~\ref{fig:PV_visc0} in terms of the pressure-volume diagram for the magnetic detonation structure similar to Fig. \ref{fig:Hugoniot_1}. We can see immediately that, since magnetic detonation is extremely weak, the shock curve (blue) and the detonation curve (red) almost coincide; similar to Fig. \ref{fig:Hugoniot_1}, the tangent line from the initial state to the CJ detonation regime is shown by green. In the case of zero viscosity, pressure at the tangent line (label "$t$") follows from Eqs. (\ref{eq:mass1h}), (\ref{eq:momentum1h}) as
\begin{equation}
\label{eq:tang}
P_t=P_0+ \rho_0 D^2\left(1-1/r\right).
\end{equation}
In order to make difference between the shock and detonation visible, we modify the pressure-volume diagram by extracting pressure at the tangent line,
Eq. (\ref{eq:tang}), i.e. plotting $P-P_{t}$ versus $V$, see Fig.~\ref{fig:PV_kappa}. Within such a representation, the tangent line (by definition) becomes simply the zero-line, while the shock and the detonation states may be well-distinguished and represented by two curves resembling parabola pieces.


\subsection{The analytical theory for the key magnetic detonation parameters}
As we can see in Fig.~\ref{fig:PV_kappa}, the limit of weak compression $\left( r-1\right)\ll1$ holds with a very good accuracy for magnetic detonations, and hence allows developing the analytical theory of the process.
The system Eqs. (\ref{eq:mass1h})-(\ref{eq:energy1h}) may be reduced to a single equation, which is equivalent to the Hugoniot relation\cite{LL_VI} with thermal conduction taken into account
\begin{equation}
\label{eq:rel_2}
\varepsilon-\varepsilon_0-\frac{1}{2}\frac{r-1}{r}\frac{P_0+P}{\rho_0}+Q(a-1)=\frac{\kappa}{\rho_0 D} \frac{dT}{dz}.
\end{equation}
By combining the equations of state for the crystals of nanomagnets, Eqs. (\ref{eq:presure1}), (\ref{eq:st_energy1}), we get rid of the thermal terms and obtain the expression for thermodynamic energy as function of $P$ and $r$ only
\begin{align}
\label{eq:rel_3}
\varepsilon=&\frac{1}{r\Gamma}\frac{P}{\rho_0} \nonumber \\
&+\frac{c_0^2}{nr}\left[(r^n-1)\left(\frac{1}{n-1}-\frac{1}{\Gamma}\right)-\frac{n}{n-1}(r-1)\right].
\end{align}
Then substituting Eq. (\ref{eq:rel_3}) into the Hugonoit relation Eq. (\ref{eq:rel_2}) we obtain pressure as
\begin{align}
\label{eq:rel_4}
\frac{P}{\rho_0}=&\frac{2\Gamma}{2-\Gamma(1-r)} \left\{ rQ(1-a)  \right.\nonumber \\
&+\left( \frac{r-1}{2}\Gamma+r\right)\varepsilon_0+r\frac{\kappa}{\rho_0 D}\frac{dT}{dz} \nonumber \\
&\left. +\frac{c_0^2}{n}\left[\frac{n}{n-1}(r-1) -(r^n-1)\left(\frac{1}{n-1}-\frac{1}{\Gamma}\right)   \right]  \right\}.
\end{align}
In a similar way, by substituting Eqs. (\ref{eq:presure1}), (\ref{eq:st_energy1}) into Eq. (\ref{eq:rel_2}) we find temperature
\begin{align}
\label{eq:rel_3}
T^{\alpha+1}=&\frac{2(\alpha+1)M\Theta_D^{\alpha}}{RA\left[2-\Gamma(1-r)\right]} \left\{ Q(1-a)  \right.\nonumber \\
&\left. +\frac{c_0^2}{nr}\left[\frac{n}{n-1}(r-1) - \left( \frac{1}{n-1} -\frac{r-1}{2}\right)    \right]   \right. \nonumber \\
&\left.  +\left( \frac{r-1}{2r}\Gamma+1\right)\varepsilon_0+\frac{\kappa}{\rho_0 D}\frac{dT}{dz} \right\}.
\end{align}

Now we take into account that density variations in the magnetic detonation are small by introducing a small value $\delta = r -1 =\rho/\rho_0 -1\ll1$. We can also neglect initial thermodynamic energy since $\varepsilon_0/Q\sim 10^{-6}$.
First, we analyze the detonation products, $z\rightarrow\infty$, for which the transformation process is completed, $dT/dz\rightarrow0$.
Expanding Eq. (\ref{eq:rel_4}) in powers of $\delta$ and taking $a=0$ we obtain pressure in the  detonation products as
\begin{align}
\label{eq:rel_6}
\frac{P_d}{\rho_0}=&\Gamma Q  +\left[c_0^2+\Gamma \left(\frac{\Gamma}{2}+1\right)\right]\delta_{d} \nonumber \\ &+\frac{1}{2}\left[\frac{\Gamma^2}{2}Q\left(\frac{\Gamma}{2}+1\right)+c_0^2(n-1)\right]\delta_{d}^2.
\end{align}
The scaled density deviation $\delta_d$ corresponding to the CJ point of the detonation products on Fig. \ref{fig:PV_kappa} is unknown so far and we have to find it. At the CJ point the tangent line touches  the detonation curve, which implies
\begin{align}
\label{tang_cond1}
P_t=&P_d,
\end{align}
\begin{align}
\label{tang_cond2}
\frac{\partial P_t}{\partial r}=&\frac{\partial P_d}{\partial r}.
\end{align}
For small compression case $\delta\ll1$ we rewrite Eq. (\ref{eq:tang}) as
\begin{equation}
\label{eq:tang3}
\frac{P_t-P_{0}}{\rho_0}=D^2 \delta(1-\delta),
\end{equation}
Substituting Eqs.(\ref{eq:rel_6}) and (\ref{eq:tang3}) into Eq.(\ref{tang_cond1}), (\ref{tang_cond2}), we find
\begin{align}
\label{eq:sys1}
D^2\delta_d(1-\delta_d)=&\Gamma Q  +\left[c_0^2+\Gamma Q\left(\frac{\Gamma}{2}+1\right)\right]\delta_d\nonumber \\ &+\frac{1}{2}\left[\frac{\Gamma^2}{2}Q\left(\frac{\Gamma}{2}+1\right)+c_0^2(n-1)\right]\delta_d^2,
\end{align}
\begin{align}
\label{eq:sys2}
D^2(1-2\delta_d)=&c_0^2+\Gamma Q \left(\frac{\Gamma}{2}+1\right)\nonumber \\
&+\left[\frac{\Gamma^2}{2}Q\left(\frac{\Gamma}{2}+1\right)+c_0^2(n-1)\right]\delta_d.
\end{align}
Then, eliminating $D^{2}$ from Eqs. (\ref{eq:sys1}), (\ref{eq:sys2}), and keeping terms as small as $\propto \delta_d^{2}$ we obtain
\begin{align}
\label{eq:rel_7}
\Gamma Q\left(1-2\Gamma\delta_d\right)=\nonumber \\
+&\delta_d^2\left[\Gamma Q \left(1+\frac{\Gamma}{2}\right)\left(1+\frac{\Gamma}{4}\right)+\frac{1}{2}c_0^2(n+1)\right]
\end{align}
We may also use the condition $4 \Gamma^2 Q \ll \delta_d (n+1)c_0^2$ justified below to simplify Eq. (\ref{eq:rel_7}) and find the density deviations in the detonation products as
\begin{equation}
\label{eq:delta_d}
\delta_d=\frac{1}{c_0}\sqrt{\frac{2\Gamma Q}{n+1}}.
\end{equation}
As an example, for the external magnetic field of $B_z=4\textrm{T}$ we obtain the density deviations from the initial value as small as $\delta_d\approx9.6\cdot10^{-3} \ll 1$.
Temperature of the detonation products may be calculated with the accuracy of the first order terms in $Q \delta_{d}$ as
\begin{align}
\label{eq:rel_6T}
T_d^{\alpha+1}=\frac{(\alpha+1)M\Theta_D^{\alpha} }{RA}\left[ Q\left(1+\frac{\Gamma}{2}\delta_{d}\right)+c_{0}^{2}\frac{n+1}{12} \delta_{d}^{3}\right] =\nonumber \\
\frac{(\alpha+1)QM\Theta_D^{\alpha} }{RA}\left(1+\frac{7\Gamma}{6c_0} \sqrt{\frac{\Gamma Q}{2(n+1)}}\right).
\end{align}
\begin{figure}
\includegraphics[width=3.4in]{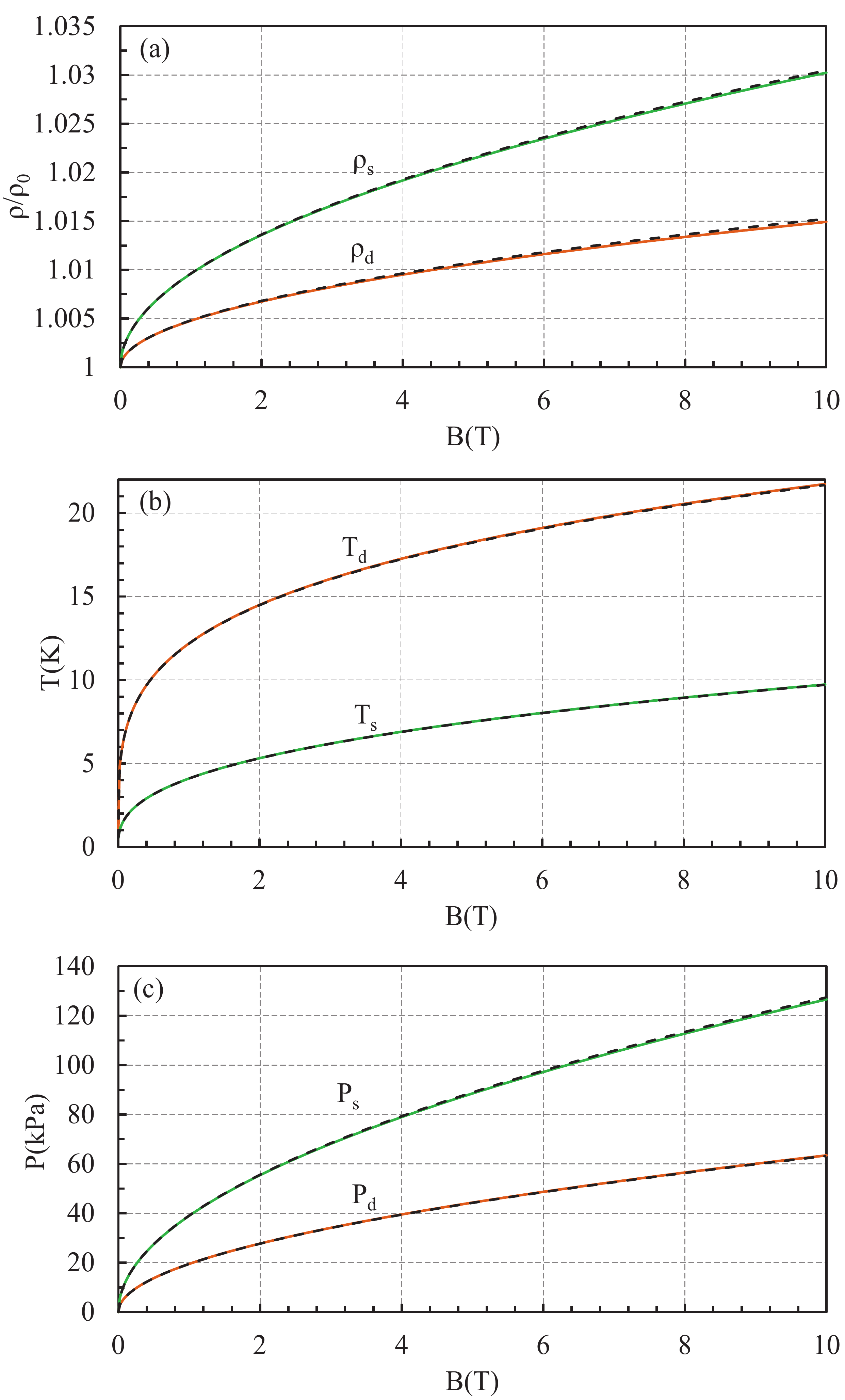}
\caption{Density, temperature and pressure at the leading shock and in the magnetic detonation products versus the external magnetic field. Solid lines correspond to the numerical solution and the dashed lines represent the analytical theory.}
\label{par_vs_B}
\end{figure}
Figure \ref{par_vs_B} shows density, temperature and pressure of the detonation products versus the applied magnetic field as found in the numerical solution and predicted by the analytical formulas Eq. (\ref{eq:rel_6}), (\ref{eq:delta_d}), (\ref{eq:rel_6T}); we observe excellent agreement of the theory with the numerical solution. We also point out that the magnetic detonation speed $D^{2}$ may be obtained
 from  Eq. (\ref{eq:sys2}) up to the leading terms in $\delta_{d}$ as:
\begin{align}
\label{eq:D}
D^{2}=&\frac{1}{1-2\delta_d}\left\{c_0^2+\Gamma Q \left(\frac{\Gamma}{2}+1\right)\right. \nonumber \\
&\left.+\left[\frac{\Gamma^2}{2}Q\left(\frac{\Gamma}{2}+1\right)+c_0^2(n-1)\right]\delta_d\right\},
\end{align}
Keeping the leading terms we can rewrite Eq.(\ref{eq:D}) as:
\begin{align}
\label{eq:D2}
D\approx c_0+\frac{\Gamma Q}{2c_0^2}\left(\frac{\Gamma}{2}+1\right)+\frac{\delta_d}{2}\left[\frac{\Gamma Q}{2c_0^2} {\left(\frac{\Gamma}{2}+1\right)}^{2}+(n+1)\right]
\end{align}
Substituting Eq. (\ref{eq:delta_d}) in Eq. (\ref{eq:D2}) and taking into account $\Gamma Q/c_0^2\ll 1$ we find the magnetic detonation speed
\begin{align}
\label{eq:D3}
D\approx c_0+\sqrt{\frac{n+1}{2}\Gamma Q}
\end{align}
Thus, in agreement with Ref. [\onlinecite{Modestov-det}], the magnetic detonation speed only slightly exceeds the sound speed in the crystals.

It is also useful finding thermodynamic parameters of the crystal just behind the shock, $a=1$, neglecting thermal conduction, i.e. treating the  shock as a discontinuity similar to  Ref. [\onlinecite{Modestov-det}]. We will show in the next subsection, that the same values for temperature, pressure and density hold at the isothermal discontinuity taking into account thermal conduction.
So, we are looking for the density deviations $\delta_{s}$ corresponding to the crystal just behind an infinitely thin shock.
Equation (\ref{eq:rel_4}) yields up to the second order terms in $\delta$
\begin{align}
\label{eq:rel_10}
\frac{P_s}{\rho_0}=\delta c_0^2 \left[1 +\frac{1}{2}\delta (n-1)\right].
\end{align}
The tangent line intersects the  shock curve at the point $s$, which implies $P_s=P_t$ or
\begin{align}
\label{eq:rel_8}
D^{2}\delta_s(1-\delta_s)=\delta_{s} c_0^2 \left[1 +\frac{1}{2}\delta_{s} (n-1)\right]
\end{align}
Taking into account the result for the magnetic detonation speed, Eq. (\ref{eq:D}), we obtain the equation for $\delta_s$
\begin{align}
\label{eq:rel_11}
\left(1-\delta_s\right)\left\{   c_0^2+\left[\frac{\Gamma^2}{2}Q\left(\frac{\Gamma}{2}+1\right)+c_0^2(n-1)\right]\delta_d \right. \nonumber\\
\left. +\Gamma Q \left(\frac{\Gamma}{2}+1\right) \right\}=
\left(1-2\delta_d\right)c_0^2 \left[1 +\frac{1}{2}\delta_s (n-1)\right],
\end{align}
which may be solved with the accuracy of $\Gamma Q\ll c_0^2 \delta$ as
\begin{align}
\label{eq:delta_s2}
\delta_s\approx 2\delta_d
\end{align}
It is noted that density deviation just behind the discontinuous shock is about twice larger than in the detonation products, which may be seen already in Fig.~\ref{fig:PV_kappa}.
Then we find temperature on the shock similar to Eq. (\ref{eq:rel_6T})
\begin{align}
\label{eq:rel_7T}
T_s^{\alpha+1}=\frac{(\alpha+1)M\Theta_D^{\alpha} }{RA}\frac{n+1}{12} c_{0}\delta_{s}^{3} \approx\nonumber \\
\frac{4(\alpha+1)\Gamma QM\Theta_D^{\alpha} }{3 c_0 RA} \sqrt{\frac{2 \Gamma Q}{n+1}}.
\end{align}
Density, temperature and pressure at the leading shock in magnetic detonation are shown in Fig. \ref{par_vs_B}; we again observe a very good agreement of the analytical theory and the numerical solution.

\subsection{Isothermal discontinuity in magnetic detonation}
\begin{figure}
\includegraphics[width=3.4in]{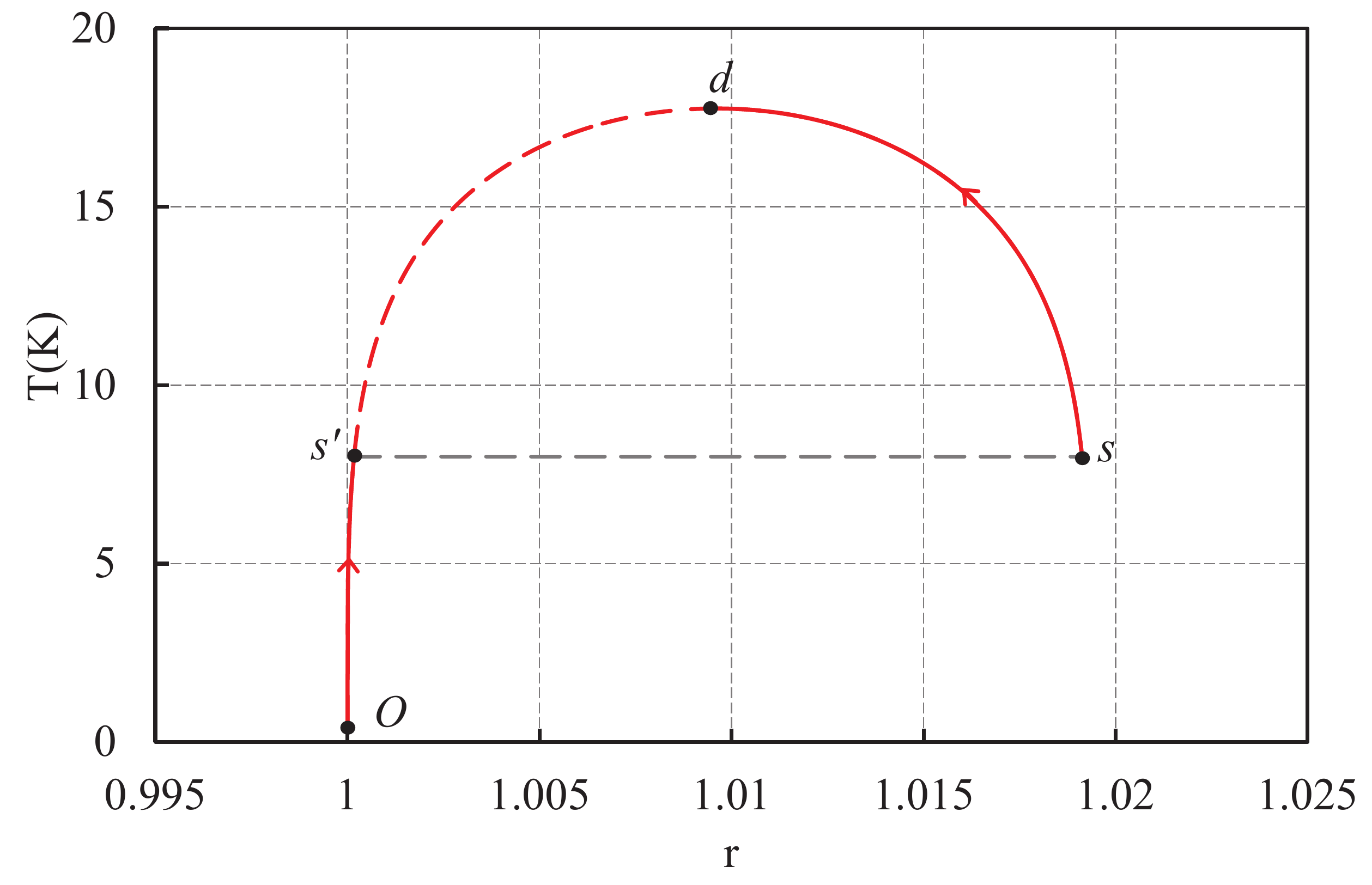}
\caption{The isothermal discontinuity diagram. Temperature dependence on scaled density $r$ according to Eq.(\ref{eq:temp_kappa}). Starting from the initial point $O$, all parameters vary continuously to $s'$. Then pressure and density experience isothermal discontinuity from the point $s'$ to the point $s$ (the black dashed line). The red dashed line corresponds to the nonphysical region in the shock phase. From the point $s$, the system relaxes to the final state $d$ in the process of spin-flipping (magnetic detonation).
Crystal compression at the point $s'$  is $r=1.00012$.}
\label{fig:tempjump}
\end{figure}

In this subsection we obtain the magnetic detonation structure in presence of thermal conduction.
By using Eqs. (\ref{eq:presure1}), (\ref{eq:tang}) we find temperature as a function of scaled density $r$ inside the magnetic detonation front
\begin{equation}
\label{eq:temp_kappa}
T^{\alpha+1}= \frac{(\alpha +1)M \Theta_{D}^{\alpha}}{\Gamma R A r}\left[D^2\left(1-\frac{1}{r}\right)-
\frac{c_0^2}{n}\left(r^n-1\right)\right],
\end{equation}
By plotting temperature $T$ according to Eq. (\ref{eq:temp_kappa}) from the initial point $O$ to the shock point $s$, we observe non-monotonic temperature variations, see Fig.~\ref{fig:tempjump}, with the maximum value attained at the detonation point $d$. In that case, assuming hypothetic temperature variations along the red curve (solid and dashed) in Fig.~\ref{fig:tempjump} from $O$ to $d$ in the whole detonation front, one moves first from $O$ to $s'$, then passes $d$ to the shock point $s$, and then returns back from $s$ to the final state $d$. Along such a pass we obtain the region with temperature decrease $dT/dx<0$ on the way from $d$ to the shock point $s$.
  However, the left hand-side of Eq. (\ref{eq:rel_2}) turns to $0$ only at the boundaries of the detonation wave $z\rightarrow\pm\infty$, i.e. it has to be either positive or negative in the transition zone from $0$ to $d$, see Refs. [\onlinecite{Zeldovich-02,LL_VI}]. Since detonation definitely leads to temperature increase as compared to the initial state, then we have the condition $dT/dx>0$ satisfied everywhere in the magnetic detonation front, so that the hypothetic region with $dT/dx<0$ is not physical and should be avoided.
 Then, variations of all values in magnetic detonation in Fig.~\ref{fig:tempjump} have to correspond to continuous movement from the initial point $O$ to the point $s'$, which is followed by jump to the shock point $s$ with subsequent continuous evolution to the final detonation point $d$. At the same time, because of thermal conduction in Eqs. (\ref{eq:mass1h}) -  (\ref{eq:energy1h}), temperature has to be continuous in magnetic  detonation including the shock wave.
For that reason, instead of a traditional shock, we obtain isothermal discontinuity in magnetic detonation in the case of negligible volume viscosity when the shock structure is supported by thermal conduction only. An important point is that compression of the crystal in the point $s'$ is quite small, $r=1.00012$, so that the isothermal discontinuity resembles a shock for density and pressure. The visual difference is found only for temperature variations.

\begin{figure}
\includegraphics[width=3.4in]{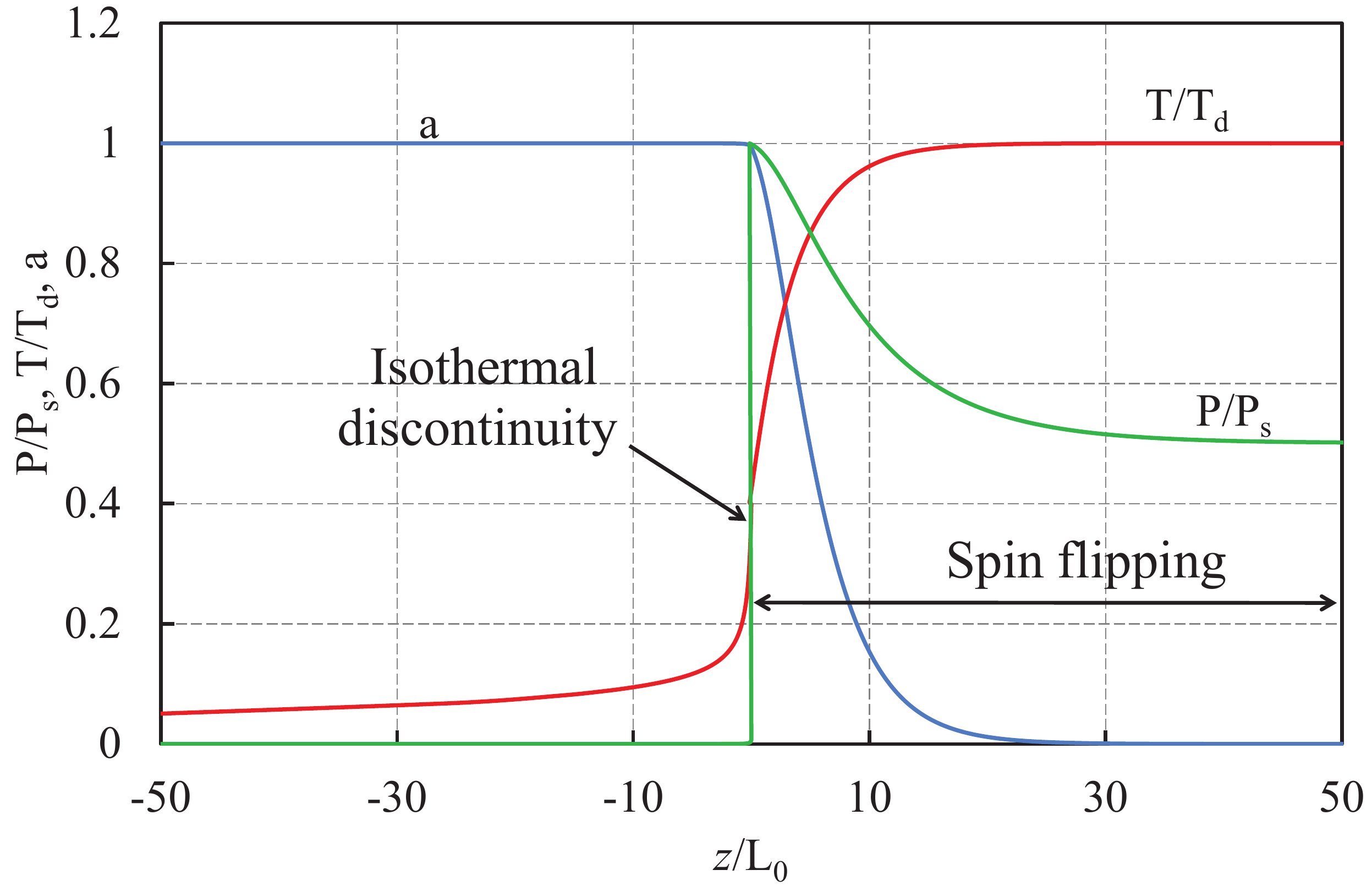}
\caption{\label{fig:Profile_kappa}
Magnetic detonation structure in Mn$_{12}$-acetate crystals in the external magnetic field $H=4\,\mathrm{T}$. The plots are scaled by the temperature of the detonation products $T_d=17.25\,\mathrm{K}$ and the shock pressure $P_s\approx80\,\mathrm{kPa}$. Other parameters are specified in the text.}
\end{figure}

The internal structure  of  magnetic detonation can be obtained by integrating the equation for kinetics of spin-relaxation,\cite{Garanin-Chudnovsky-2007} which resembles strongly the Arrhenius law of chemical kinetics Eq. (\ref{burning_1}),
\begin{equation}
\label{eq:Arrhenius2}
\frac{\partial a}{\partial t}=-\frac{a}{\tau} \exp(-E_{a}/T),
\end{equation}
By adopting the reference frame of the stationary magnetic detonation front, Eq. (\ref{eq:Arrhenius2}) can be written as
\begin{equation}
\label{eq:Arrhenius3}
u\frac{\partial a}{\partial x}=-\frac{a}{\tau} \exp(-E_{a}/T).
\end{equation}
By using Eq. (\ref{eq:temp_kappa}),  we solve Eq.(\ref{eq:Arrhenius3}) numerically starting from the initial point $O$  to the point of isothermal discontinuity $s'$. After that new values for density and pressure at the point $s$ are calculated according to the jump conditions, with temperature and fraction of metastable molecules $a$ kept constant.
Then numerical calculation is performed from the point  $s$ to the final detonation state $d$. The result of the numerical solution is presented in Fig.~\ref{fig:Profile_kappa} for the Mn$_{12}$-acetate crystals with the characteristic thickness  of the detonation front $L_0\equiv c_0 \tau\approx0.2\,\mathrm{mm}$ employed as the length scale.
In Fig.~\ref{fig:Profile_kappa}  for illustrative purposes we plot the numerical solution for the thermal conduction coefficient $\kappa=10^4\, \textrm{m}^2/\textrm{s} $, which is about five times larger than the commonly accepted value $\kappa\approx 2\cdot10^{-5}\,\textrm{m}^2/\textrm{s}$, e.g. see Refs. [\onlinecite{Garanin-Chudnovsky-2007,Dion-2013}]. Still, even in that case thickness of the heating region is much smaller than the region of spin-flipping, and fraction of the molecules in the metastable state is close to unity at the isothermal discontinuity, $a=0.993$.

Thus, as an intermediate conclusion, thermal conduction influences only the temperature profile in magnetic detonation, with a minor effect on density and pressure, and with negligible modifications of the  total front thickness. Much more dramatic modifications of the magnetic detonation structure are expected because of volume viscosity as shown in the next section.

\section{Detonation front with viscosity}
By taking into account volume viscosity, $\eta$, we re-write the
equations of  mass, momentum and energy conservation for magnetic detonation as
\begin{equation}
\label{eq:mass1v}
\rho_0 u_0=\rho u,
\end{equation}
\begin{equation}
\label{eq:momentum1v}
P_0+\rho_0 u_0^2=P+\rho u^2-\eta \frac{du}{dx},
\end{equation}
\begin{align}
\label{eq:energy1v}
\rho_0 u_0 & \left( \epsilon_0+\frac{P_0}{\rho_0}+\frac{u^{2}}{2}\right)=\nonumber \\
&\rho u \left( \epsilon+\frac{P}{\rho}+\frac{u^{2}}{2}\right)-\kappa\frac{dT}{dx}-\eta u \frac{du}{dx}.
\end{align}
We point out that, unlike commonly known shear viscosity arising due to relative motion of gas or fluid layers, volume viscosity describes momentum and energy dissipations due to compression of a medium. While shear viscosity is not typical for solid state processes such as magnetic detonation in crystals of nanomagnets, volume viscosity has to be considered.
It is convenient characterizing the role of viscosity by the  dimensionless parameter
\begin{equation}
\label{eq:dimensional-visc}
\eta'\equiv\frac{\eta}{\rho_0 c_0 L_0}=\frac{\eta}{\rho_0 c_0^{2} \tau}.
\end{equation}
This dimensionless viscosity plays a  role conceptually similar to that of the inverse Reynolds number in fluid mechanics.\cite{LL_VI}
In particular, the parameter values $\eta'=0.0025; \, 0.005; \, 0.01$ employed below correspond to the domain of the Reynolds numbers $\textrm{Re} = 100 - 400$. At such values of the Reynolds number gas and fluid flows are typically laminar; negligible role of viscosity is qualitatively indicated by transition to turbulence, which happens usually at larger values, $\textrm{Re} \sim 10^{3}$ and above.
To the best of our knowledge, there have been no works, either experimental or theoretical, investigating  volume viscosity in crystals of nanomagnets, and therefore we will take $\eta'$ as a free parameter.

\begin{figure}
\includegraphics[width=3.4in]{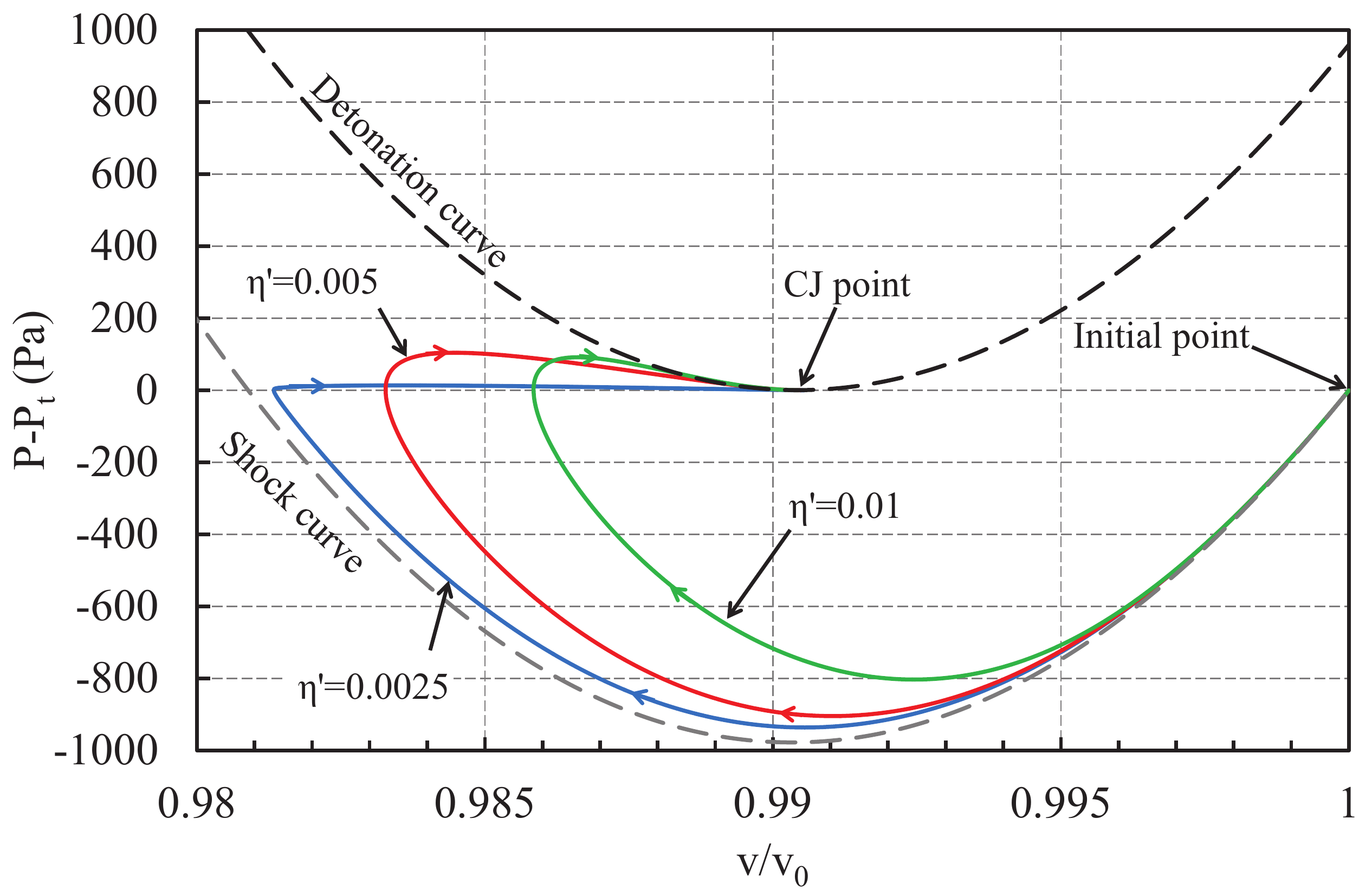}
\caption{\label{fig:PV-diagram-viscosity} The pressure-volume diagram for different values of the scaled viscosity $\eta'=0.0025; \, 0.005; \, 0.01$ in the external magnetic field $B_z=4\textrm{T}$.}
\end{figure}

\begin{figure}
\includegraphics[width=3.4in]{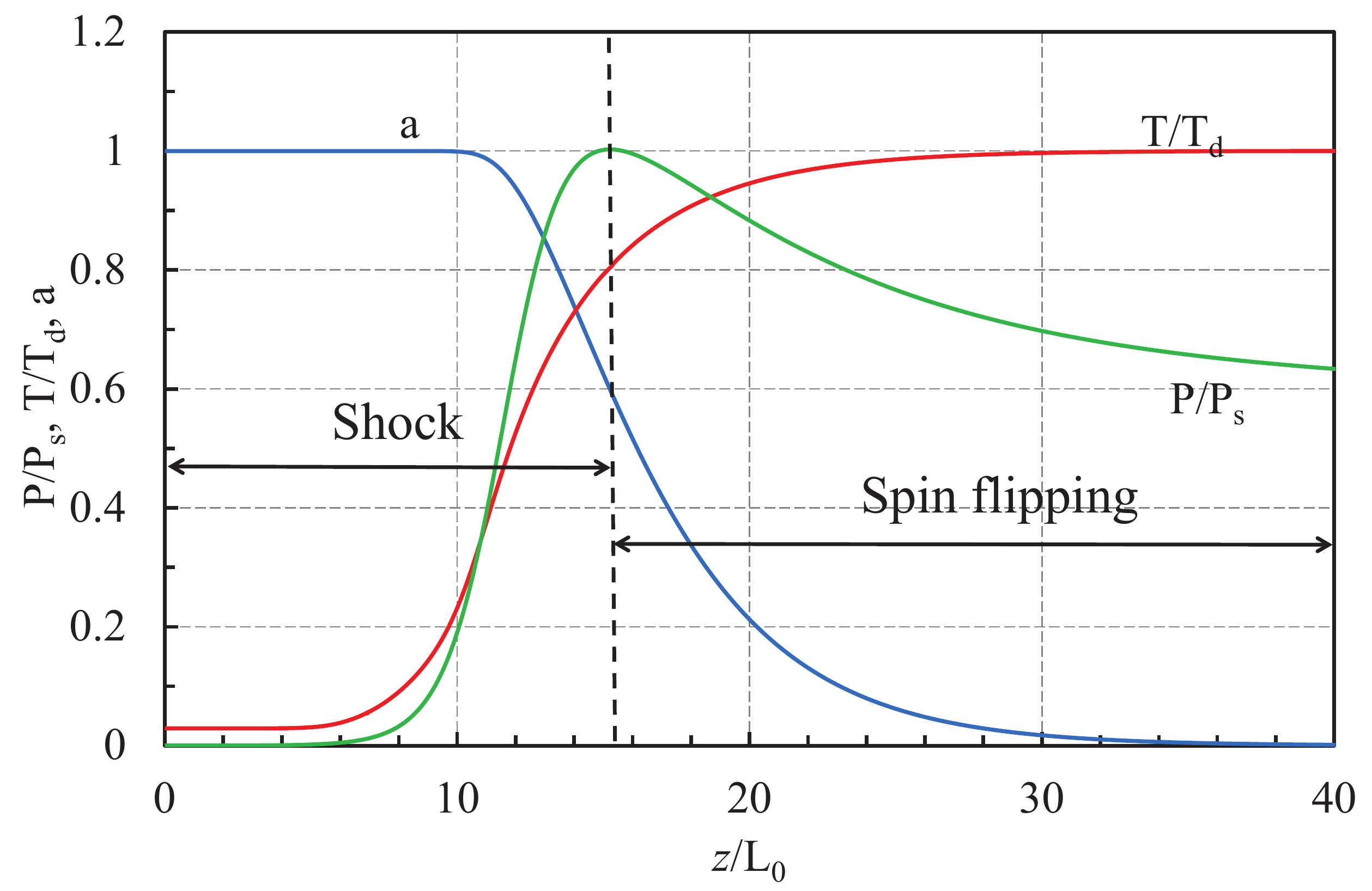}
\caption{\label{fig:profile_visc} Profiles of scaled pressure, temperature and fraction of nanomagnets in the metastable state  for magnetic detonation for  scaled viscosity $\eta'=0.005$ in the external magnetic field $B_z=4\textrm{T}$.}
\end{figure}

\begin{figure}
\includegraphics[width=3.4in]{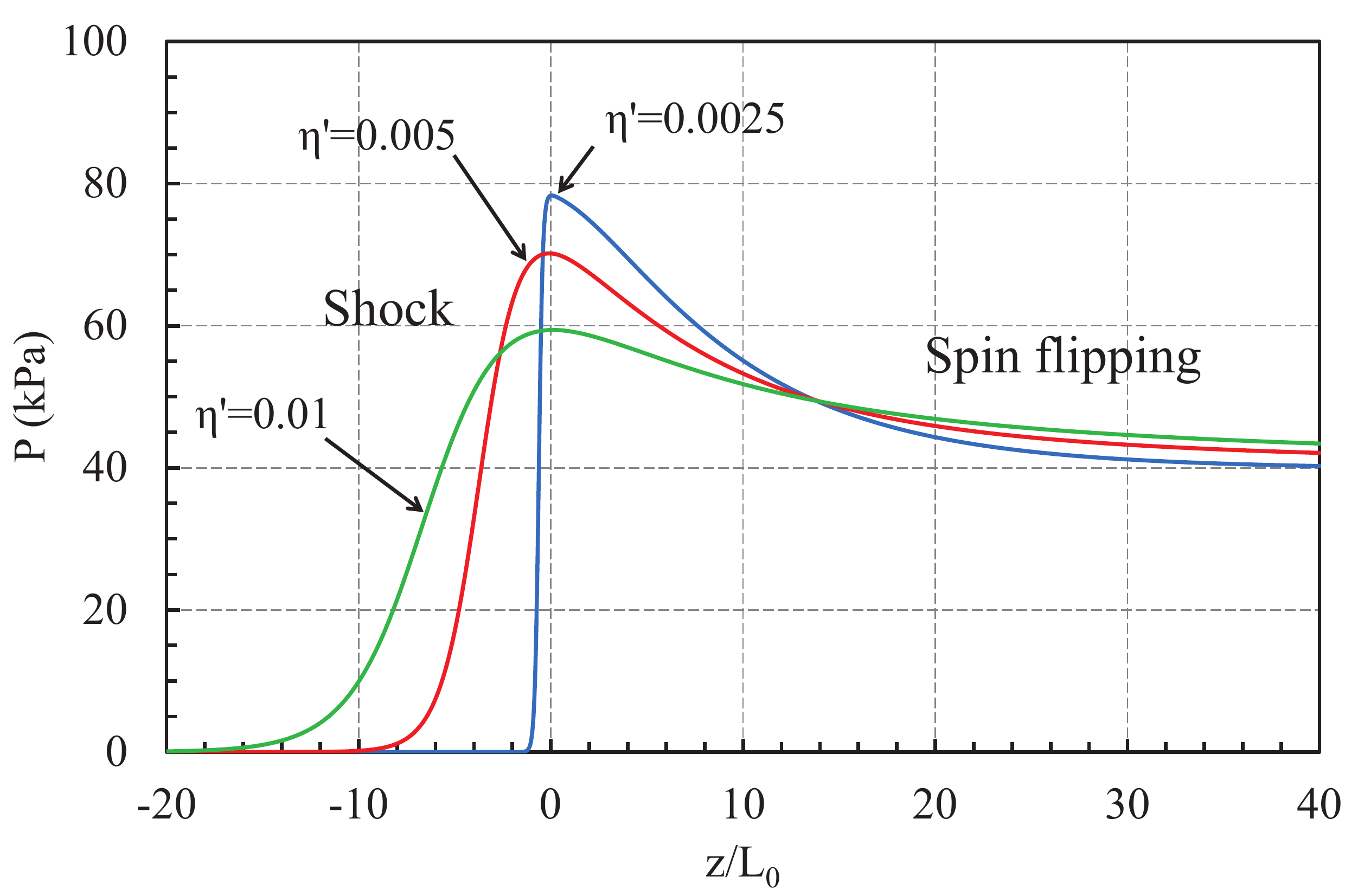}
\caption{\label{fig:profile_visc_p} Pressure profiles in magnetic detonation for different values of the scaled viscosity $\eta'=0.0025; \, 0.005; \, 0.01$ in the external magnetic field $B_z=4\textrm{T}$. Pressure maxima are placed at the position $z=0$, which may be defined as the back side of the shock region in the detonation.}
\end{figure}

\begin{figure}
\includegraphics[width=3.4in]{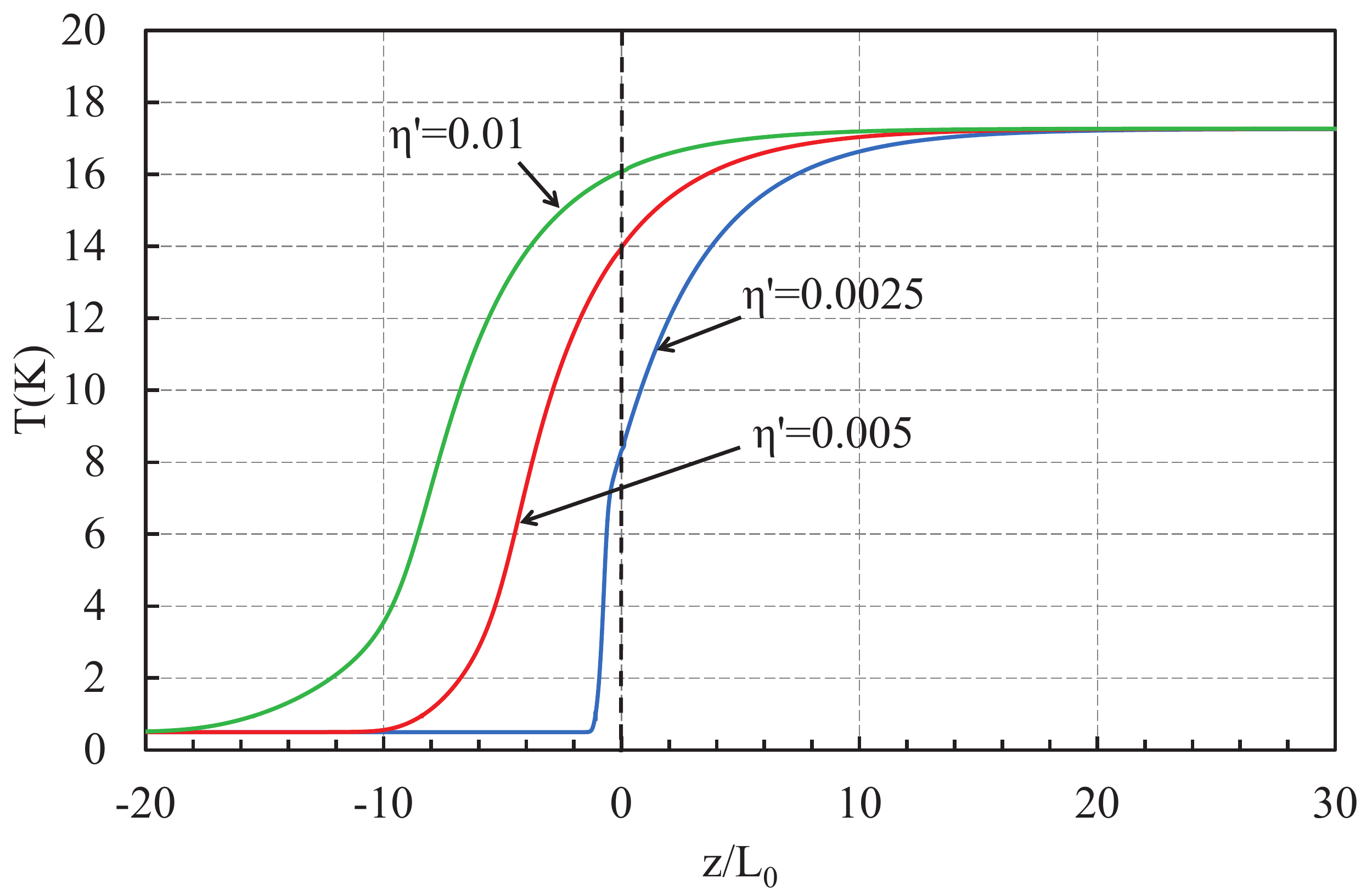}
\caption{\label{fig:profile_visc_T} Temperature profiles in magnetic detonation for different values of the scaled viscosity $\eta'=0.0025; \, 0.005; \, 0.01$ in the external magnetic field $B_z=4\textrm{T}$. Similar to Fig.~\ref{fig:profile_visc_p}, position $z=0$ corresponds to the maximum pressure for each profile treated as the back side of the shock wave.}
\end{figure}

We stress that even small values of volume viscosity  $\eta' \sim 0.01$ lead to considerable changes in the magnetic detonation structure as shown in Figs.~\ref{fig:PV-diagram-viscosity} - \ref{fig:profile_visc_T}.  Even on the pressure-volume diagram, Fig.~\ref{fig:PV-diagram-viscosity}, (plotted, as before, for $P - P_{t}$ by extracting the CJ-tangent line for illustrative purposes) we observe that all discontinuous jumps of the previous model\cite{Modestov-det} are replaced by continuous transition lines from the initial point to the CJ detonation products. For very small viscosity, $\eta'=0.0025$, the transition line (blue), although continuous, goes pretty close to the Hugoniot curve for the shock and then to the tangent line for the spin-flipping process. In a similar way, in Fig.~\ref{fig:profile_visc_p} the plot for $\eta'=0.0025$ demonstrates quite abrupt initial pressure increase -- almost jump, which resembles strongly a discontinuous shock wave; this pressure increase is followed by pressure relaxation to a smaller value in the process of spin-flipping with Zeeman energy release. As we take larger values of volume viscosity, $\eta'= 0.005; \, 0.01$, deviation of the pressure-volume plots from the discontinuous shock model becomes much more pronounced, Fig.~\ref{fig:PV-diagram-viscosity}, and we find pressure and density in the shock wave decreasing as compared to the discontinuous case, Fig.~\ref{fig:profile_visc_p}. In fact, the very definition of a shock as a part of magnetic detonation front becomes ambiguous when volume viscosity is taken into account. To avoid the ambiguity, we notice that within the model of a discontinuous shock, pressure maximum is attained at the shock front, and then pressure goes down as the spin-flipping starts. In the same way it seems natural  treating the point of maximal pressure in Figs.~\ref{fig:profile_visc}, \ref{fig:profile_visc_p} as the back-side of the shock region. In Fig.~\ref{fig:profile_visc_p} we placed all three plots by choosing $z=0$ as the position of the pressure maximum and hence the back side of the shocks. Then the regions corresponding to $z<0$ belong to the shocks smoothed by viscosity, while the domain of $z>0$ may be treated roughly as the regions of spin-flipping with Zeeman energy release. 
Of course, such separation is rather qualitative than quantitative, because in presence of considerable volume viscosity spin-flipping starts already in the shock, see Fig.~\ref{fig:profile_visc} corresponding to $\eta'=0.005$, $B_z=4\textrm{T}$. In Fig.~\ref{fig:profile_visc} the fraction of nanomagnets in the metastable state is about $a \approx 0.6$ at the point of pressure maximum. Figure~\ref{fig:profile_visc_T} compares temperature profiles for the scaled viscosity $\eta'= 0.0025; \, 0.005; \, 0.01$ with the respective pressure maxima still placed at $z=0$. As we can see, larger volume viscosity increases strongly temperature at the shock; in the case of $\eta'= 0.01$ temperature at $z=0$ is only 5\% smaller than the final detonation temperature, i.e. it is about $\sim 0.95 T_{d}$. Temperature increase reduces strongly the size of the "pure" spin-flipping region, still, as we can see from  Fig.~\ref{fig:profile_visc_T}, it is accompanied by strong increase of the width of the shock wave, so that total width of magnetic detonation front does not change much, being about $\sim 10 L_{0}$ for all three cases, i.e. about $1-2$ mm in dimensional units.

We point out that the experimentally employed sample sizes for the crystals of nanomagnets are also about 2 mm, e.g. see Refs. [\onlinecite{Decelle-09,Subedi-2013}]. As a result, it is rather difficult to observe steady well-developed magnetic detonation in the commonly experimental conditions. Instead, we suggest that most of the experimental points  reported in Ref. [\onlinecite{Decelle-09}] for ultra-fast magnetic avalanches correspond to magnetic detonation in the process of development, which is also indicated by the average avalanche speed in the samples noticeably below the sound speed. Another important feature of the experimental observations of Ref. [\onlinecite{Decelle-09}] is that the ultra-fast avalanches were obtained for the magnetic field close to the quantum resonance values of the nanomagnets. Quantum resonances lead to strong decrease of the factor $\tau$ in the kinetic equation of spin-flipping, Eq. (\ref{eq:Arrhenius2}), and hence of the characteristic width of the magnetic detonation front $\propto L_{0}=c_{0}\tau$, which may allow experimental observation of magnetic detonation in samples of conventionally employed sizes.

\section{Summary}
In the present paper we have investigated the internal structure of magnetic detonation in crystals of nanomagnets. Magnetic detonation is weak and propagates with speed only slightly exceeding the sound speed in the crystals. For that reason, in stark contrast to usual combustion detonations, transport processes -- thermal conduction and volume viscosity -- play an important role in forming the magnetic detonation structure. We show that, in the case of negligible volume viscosity, thermal conduction produces isothermal discontinuity instead of the leading shock in magnetic detonation. In the isothermal discontinuity temperature of the crystal is continuous, while density and pressure experience jump.

Volume viscosity leads to much more dramatic changes of the magnetic detonation structure as compared to the model of Ref. [\onlinecite{Modestov-det}] with neglected transport processes. In that case all important thermodynamic parameters of the crystals acquire smooth profiles all over the magnetic detonation front including the leading shock. In addition, the very concept of the leading shock requires unambiguous definition -- here we suggest to specify the back side of the shock as the position of the pressure maximum. As the relative role of volume viscosity increases, the leading shock becomes wider and may exceed considerably  the zone of spin-flipping by  size.  Still, total size of a magnetic detonation front does not change much with variations of volume viscosity since decrease of the spin-flipping region is compensated by increase of the shock width.

\textbf{Acknowledgements}

This work has been supported by the Swedish Research Council (VR).

\end{document}